\documentclass[%
 reprint,
 amsmath,amssymb,
aip
]{revtex4-1}

\usepackage{graphicx}
\usepackage{dcolumn}
\usepackage{bm}

\begin{document}

\preprint{AIP/123-QED}

\title{On microscopic entropy production, heat and fluctuation theorem}

\author{Jianzhong Wu}
\email{jwu@engr.ucr.edu}
\affiliation{%
Department of Chemical and Environmental Engineering \& Department of Mathematics, University of California, Riverside, CA 92521, U.S.A.}%

\date{\today}

\begin{abstract}
We demonstrate that the Gibbs-Shannon entropy is applicable to non-equilibrium systems of any size and boundary conditions. The change in microscopic entropy can be attributed to the stochastic nature of dynamic processes and to the inherent uncertainties of thermodynamic systems. The latter predicts that entropy production is nonnegative on average and varies with different trajectories according to the fluctuation theorem. By contrast, heat is affiliated with stochastic processes underlying particle motions and the ensemble average over all possible trajectories leads to the Clausius inequality. The Jarzynski/Crooks equations can be readily derived by applying the fluctuation theorem to heat variation over different trajectories linking equilibrium states. \\

\end{abstract}

\maketitle

The universal applicability of thermodynamics is rarely questioned. However, the precise meanings of basic thermodynamic quantities like entropy, heat and work remain ambiguous from a microscopic perspective \cite{RN1,RN2, RN23, RN24}. Different definitions of entropy exist since the early days of statistical mechanics and new interpretations continue to emerge, fueling perennial debates on their applications to various thermodynamic systems, large or small both classical and quantum \cite{RN13,RN15,RN17,RN22,RN26}. Are nonequilibrium properties dependent solely on the system under investigation or on an ensemble of systems prepared according to the same macroscopic prescription \cite{RN4}? Is entropy production determined by the probability of microstates or the probability of trajectories describing the evolution of the system? Can heat transfer from cold to hot spontaneously without violating the second law of thermodynamics?  Should we use the Gibbs entropy or the Gibbs volume entropy to describe the entropy increase in isolated systems \cite{RN7,RN19}? What are the relations of these Gibbs entropies to the Boltzmann entropy, the Boltzmann surface entropy, or the observational entropy \cite{Deutsch20, RN21}? Why does the time symmetry in describing particle motions break down for irreversible phenomena to make spontaneous processes in nature follow single directions \cite{Chen20}. These and many other questions related to the fundamentals of nonequilibrium thermodynamics may be addressed, at least in part, if entropy, heat and work could be unequivocally defined in terms of microscopic activities.    

Without loss of generality, consider a multi-particle system with uncertainties at the initial condition including the number of particles and their microscope degrees of freedom (e.g., positions and momenta). According to quantum mechanics, such uncertainties are unavoidable, regardless of the system size or boundary conditions. While in the following we assume, for simplicity, that the particle motions can be described by the classical  physics, we expect that the mathematical analysis is equally applicable to quantum systems in general. 

Following the convention, we define all possible microstates of the system as an ensemble. At any moment, the system in a particular microstate is described by a probability density that satisfies the normalization condition
\begin{equation}
\int d\pmb{\nu} p(\pmb{\nu})=1
\end{equation}
where $0<p(\pmb{\nu})<1$ stands for the probability density of microstate $\pmb{\nu}$. The parameter consists of a set of variables for a full description of the dynamics of all particles. 

The Gibbs-Shannon entropy is defined in terms of the probability densities of all possible microstates       
\begin{equation}
S(t) \equiv -k_B  \int d\pmb{\nu}  p(\pmb{\nu} ,t) \ln [p(\pmb{\nu} , t)\mathcal{V}] 
\end{equation}
where $-k_B$ denotes the Boltzmann constant, $p(\pmb{\nu}, t)$ stands for the probability density of microstate $\pmb{\nu}$ at time $t$, and $\mathcal{V}$ is a constant to make the probability density inside the logarithm dimensionless. As well documented, constant $\mathcal{V}$ can be fixed by imposing a self-consistent description of the entropy from classical and quantum mechanics \footnote{A discrete distribution of microstates is used to define the Gibbs entropy while the Shannon entropy is applied to a continuous distribution of microstates.}. 

Like internal energy or any other extensive thermodynamic quantities, the total entropy corresponds to an ensemble average of the microscopic entropy 
\begin{equation}
s(\pmb{\nu},t) \equiv -k_B \ln [ p(\pmb{\nu}, t) \mathcal{V} ].
\end{equation}
The microscopic entropy above is different from the stochastic or trajectory entropy in stochastic thermodynamics \cite{RN6, Broeck14}. The latter is trajectory dependent, defined by the probability of microstate $\pmb{\nu}(t)$ at time $t$ within a specific trajectory $\pmb{\nu}(\cdot)$. Because the microscopic entropy is defined in the context of \textit{an ensemble of microstates} at any moment, it is affiliated both with the uncertainties of the system introduced at the initial condition and with the stochastic nature of the particle motions (e.g., due to environmental effects). By contrast, a trajectory is defined by a possible evolution of the microstate of the system under consideration. Apparently, the trajectories would be different if the system starts from different microstates at time $t=0$. Besides, a trajectory is, in general, not uniquely determined by the initial condition (unless particle motions are deterministic). For example, a large number of trajectories are possible for a Brownian particle starting at the same initial microstate. Such uncertainties arise from the stochastic nature of the particle motion.   
 
As the system undergoes dynamic changes from $t=0$ to $t=\tau$, all microstates in the ensemble evolve accordingly. Accordingly, the change in microscopic entropy is given by
\begin{equation}
\Delta s \equiv s[\pmb{\nu}_\tau, \tau] -s[\pmb{\nu}_0,0] 
\end{equation}
where $\pmb{\nu}_0$ and $\pmb{\nu}_\tau$ represent possible microstates of the system at $t=0$ and $t=\tau$, respectively. It is important to recognize that microstates $\pmb{\nu}_0$ and $\pmb{\nu}_\tau$ may be connected by \textit{multiple trajectories}, each with its own probability density. In other words, $\Delta s$ accounts for the change in microscopic entropy due to all possible trajectories from microstate $\pmb{\nu}_0$ to $\pmb{\nu}_\tau$.   

Different microstates at the initial condition generate an ensemble of possible trajectories for the system. For each trajectory $\pmb{\nu}(\cdot)$, the microstate probability at $t=0$ can be expressed as the ratio of the trajectory probability density and the conditional probability density with the trajectory starting from microstate $\pmb{\nu}(0)=\pmb{\nu}_0$
\begin{equation}
\label{eq:traj0} 
p(\pmb{\nu}_0) = \frac{p[\pmb{\nu}(\cdot)]}{p[\pmb{\nu}(\cdot)|\pmb{\nu}_0]} 
\end{equation}
where $p(\pmb{\nu}_0)$ represents the probability density of the system in microstate $\pmb{\nu}_0$ at $t=0$. The conditional probability is defined as
\begin{equation}
p[\pmb{\nu}(\cdot)|\pmb{\nu}_0] = \int d\pmb{\nu}_0 \delta[\pmb{\nu}(0)-\pmb{\nu}_0] p[\pmb{\nu} (\cdot)] 
\end{equation}     
where $\delta[\pmb{\nu}(0)-\pmb{\nu}_0]$ denotes the Dirac delta function. Apparently, the trajectory probability density satisfies the normalization condition
\begin{equation}
\int \mathcal{D} \pmb{\nu} (\cdot) p[\pmb{\nu} (\cdot)] = 1
\end{equation}     
where integration over $\mathcal{D} \pmb{\nu}(\cdot)$ represents a summation of all possible trajectories taking place in the system over the duration $[0,\tau]$. Eq.\eqref{eq:traj0} is valid for any trajectory over this duration.
 
Given the microstate probability at $t=0$, one can in principle calculate the probability density of the microstate at $\tau$ based on the trajectory probabilities
\begin{align}
\label{eq:ensemble_pvtau} 
p(\pmb{\nu}_{\tau}) &= \int \mathcal{D} \pmb{\nu} (\cdot) \delta[\pmb{\nu}(\tau)-\pmb{\nu}_\tau] p(\pmb{\nu}_0) p[\pmb{\nu} (\cdot)|\pmb{\nu}_0]  \notag\\
&= \int \mathcal{D} \pmb{\nu} (\cdot) \delta[\pmb{\nu}(\tau)-\pmb{\nu}_\tau] p[\pmb{\nu} (\cdot)]. 
\end{align}
Eq.\eqref{eq:ensemble_pvtau} indicates that, as expected, \textit{all possible trajectories} contribute to the probability density of  the microstate at $t=\tau$. In other words, neither $p(\pmb{\nu}_0)$ nor $p(\pmb{\nu}_{\tau})$ is solely determined by any specific trajectory. These probability densities reflect the distributions of microstates at $t=0$ and $t=\tau$ over all possible microstates. 

From a microscopic perspective, the same dynamic equations can be used to describe particle motions in forward ($t>0$) and backward ($t<0$) directions. As a result, each trajectory from microstate $\pmb{\nu}_0$ at $t=0$ to microstate $\pmb{\nu}_\tau$ at $t=\tau$ can be used to define a reverse trajectory moving backward from $\pmb{\nu}_\tau$ at $t=\tau$ to $\pmb{\nu}_0$ at $t=0$. While the forward and backward trajectories may not have the same probability density, the probability density of each microstate can also be described in terms of the statistics of reverse trajectories  
\begin{equation}
\label{eq:trajR} 
p(\pmb{\nu}_\tau) =\frac {\widetilde{p}[ \widetilde{\pmb{\nu}}(\cdot)]}{\widetilde{p}[ \widetilde{\pmb{\nu}}(\cdot)|\pmb{\nu}_\tau]} 
\end{equation}   
where $\widetilde{p}[ \widetilde{\pmb{\nu}}(\cdot)]$ represents the probability density for a reverse trajectory $\widetilde{\pmb{\nu}}(\cdot)$, and $\widetilde{p}[ \widetilde{\pmb{\nu}}(\cdot)|\pmb{\nu}_\tau]$ stands for the conditional probability of the reverse trajectory starting from microstate $\pmb{\nu}_\tau$.  

According to Eqs.\eqref{eq:traj0} and \eqref{eq:trajR}, the change in microscopic entropy can be written in terms of the probability densities for \textit{any specific pair} of forward and reverse trajectories 
\begin{align}
\label{eq:twoDs}
\Delta s &=-k_B \ln \frac{p[\pmb{\nu}_\tau, \tau]}{p[\pmb{\nu}_0,0]} \notag\\
      &=-k_B \ln \frac{p[\pmb{\nu}(\cdot)|\pmb{\nu}_0]}{\widetilde{p}[\widetilde{\pmb{\nu}}(\cdot)|\pmb{\nu}_\tau]}  +k_B \ln  \frac{p[\pmb{\nu}(\cdot)]}{\widetilde{p}[\widetilde{\pmb{\nu}}(\cdot)]} 
 \end{align}
where $\widetilde{\pmb{\nu}}(\cdot)$ represents the reverse trajectory from $\pmb{\nu}_\tau$ to $\pmb{\nu}_0$, and $\widetilde{p}$ stands for its probability density. 

Eq.\eqref{eq:twoDs} reveals two contributions to the change in microscopic entropy with distinctively different meanings. The first term on the right side of Eq.\eqref{eq:twoDs} is affiliated with the stochastic nature of the forward and reverse trajectories, i.e., uncertainties due to the motions of individual particles dictated by the specific physical laws describing the system dynamics. The net effect of forward and reverse trajectories can be identified as the flow of the microscopic entropy
\begin{equation}
\Delta_h s  \equiv -k_B \ln \frac{p[\pmb{\nu}(\cdot)|\pmb{\nu}(0)]}{\widetilde{p}[\widetilde{\pmb{\nu}}(\cdot)|\pmb{\nu}(\tau)]}. 
\end{equation}
As discussed in the following, the microscopic entropy flow is related to heat transfer between the system and its environment. It disappears for deterministic processes or if there is no difference between the probability densities of the forward and reverse trajectories (e.g., at equilibrium conditions). 

The second term on the right side of Eq.\eqref{eq:twoDs} is affiliated with the statistics of the forward and reverse trajectories generated by the ensemble. This term accounts for accumulation of the net change in the instantaneous  probability density of microstates that takes place along trajectory $\pmb{\nu}(\cdot)$ over the duration $[0,\tau]$. This part of the change in microscopic entropy defines the microscopic entropy production   
\begin{equation}
\label{eq:Sgen}
\sigma [\pmb{\nu}(\cdot)] \equiv k_B \ln  \frac{p[\pmb{\nu}(\cdot)]}{\widetilde{p}[\widetilde{\pmb{\nu}}(\cdot)]}. 
\end{equation}
Eq.\eqref {eq:Sgen} is a general form of the Crooks fluctuation theorem\cite{RN12}. Unlike the change in microscopic entropy, both the microscopic entropy flow and the microscopic entropy production are trajectory dependent. While the microscopic entropy is a ``state'' property, the microscopic entropy flow and the microscopic entropy production are ``process" variables, depending on the specific trajectory connecting the initial and final microstates.    

It is straightforward to show that the Gibbs-Shannon entropy can be expressed as an average over all possible trajectories 
\begin{align}
S(t) &= -k_B  \int d\pmb{\nu}  p(\pmb{\nu} ,t) \ln [p(\pmb{\nu} , t)\mathcal{V}] \notag\\
&= -k_B   \int d\pmb{\nu} \int \mathcal{D} \pmb{\nu} (\cdot) p[\pmb{\nu} (\cdot)] \delta[\pmb{\nu}(t)-\pmb{\nu}] \ln [p(\pmb{\nu} , t)\mathcal{V}] \notag\\
&= -k_B  \int \mathcal{D} \pmb{\nu} (\cdot) p[\pmb{\nu} (\cdot)] \ln [p(\pmb{\nu}, t)\mathcal{V}]. 
\end{align}
The trajectory average is particularly useful to demonstrate the connection between the microscopic entropy and the fluctuation theorem. Based on the two contributions to the microscopic entropy change discussed above, we can readily derive the total entropy production   
\begin{align}
\label{eq:Sgen2}
\Delta_i S (t) &= \int \mathcal{D} \pmb{\nu} (\cdot) \sigma [\pmb{\nu}(\cdot)] \notag\\  
&= k_B \int \mathcal{D} \pmb{\nu} (\cdot) p[\pmb{\nu} (\cdot)] \ln \frac{p[\pmb{\nu}(\cdot)]}{\widetilde{p}[\widetilde{\pmb{\nu}}(\cdot)]} \ge 0
\end{align}
where $\ge 0$ follows Gibbs' inequality. Eq.\eqref{eq:Sgen2} predicts the second law of thermodynamics, i.e., the mean entropy production is always nonnegative. Interestingly, the nonnegative sign can be simply attributed to the alignment of forward trajectories with the direction of time increase. In other words, the origin of thermodynamic irreversibility may be simply explained in terms of the alignment of forward trajectories and the arrow of time.  

By averaging the exponent of the negative entropy production over all possible trajectories, one can readily derive the integral fluctuation theorem \cite{Evans93} 
\begin{align}
\label{eq:Sfluc}
<\exp[-\sigma/k_B]> &=  \int \mathcal{D} \pmb{\nu} (\cdot) p[\pmb{\nu} (\cdot)] \frac{\widetilde{p}[\widetilde{\pmb{\nu}}(\cdot)]}{p[\pmb{\nu}(\cdot)]} \notag\\
&= \int \mathcal{D}\widetilde{\pmb{\nu}} \widetilde{p}[\widetilde{\pmb{\nu}}(\cdot)] = 1.
\end{align}
The probability of entropy production for a specific trajectory can be evaluated from
\begin{equation}
\label{eq:spF}
p(\sigma)= \int \mathcal{D} \pmb{\nu} (\cdot) p[\pmb{\nu} (\cdot)] \delta \Big(\sigma - k_B \ln \frac{p[\pmb{\nu} (\cdot)]} {\widetilde{p}[\widetilde{\pmb{\nu}}(\cdot)]} \Big).
\end{equation}
The probability of entropy production for the reversal trajectory is
\begin{align}
\label{eq:spR}
\widetilde{p}&(-\sigma) = \int \mathcal{D} \widetilde{\pmb{\nu}} (\cdot) \widetilde{p}[\widetilde{\pmb{\nu}} (\cdot)] \delta \Big(\sigma + k_B \ln \frac{\widetilde{p}[\widetilde{\pmb{\nu}}(\cdot)]}{p[\pmb{\nu} (\cdot)]} \Big) \notag\\
 &= \int \mathcal{D} \pmb{\nu} (\cdot) p[\pmb{\nu} (\cdot)] \exp \Big( \ln \frac {\widetilde{p}[\widetilde{\pmb{\nu}} (\cdot)] } {p[\pmb{\nu} (\cdot)]}\Big) \delta \Big(\sigma - k_B \ln \frac {p[\pmb{\nu} (\cdot)]}{\widetilde{p}[\widetilde{\pmb{\nu}}(\cdot)]} \Big) \notag\\ 
&= \exp(-\sigma/k_B) p(\sigma) 
\end{align}
which leads to an alternative form of the fluctuation theorem 
\begin{equation}
\label{eq:Sfluc0}
\frac{p(\sigma)}{\widetilde{p}(-\sigma) } = e^{\sigma/k_B}. 
\end{equation}
As the probability density of microscopic entropy production satisfies
\begin{equation}
\label{eq:Sfluc1}
\int d\sigma p(\sigma) e^{-\sigma/k_B} =  \int d\sigma \widetilde{p}(-\sigma) =1,
\end{equation}
a negative value of the microscopic entropy production is not impossible but the chance diminishes exponentially with its absolute value. Although such events could take place in a small system, it is essentially nonexistent in the macroscopic limit when the exponential terms is exceedingly small.

Our discussion so far is totally generic, independent of specific dynamics of the N-particle system. If the system is in contact with a thermal bath at temperature $T$, the entropy flow can be afflicted with the heat transfer between the system and the thermal path
\begin{equation}
\label {eq:heat}
\Delta_h s = q/T.    
\end{equation}
From a microscopic perspective, Eq.\eqref{eq:heat} may serve as a definition of heat, $q \equiv T \Delta_h s $. Substituting $\sigma = \Delta s - q/T$ into Eq.\eqref{eq:Sfluc} yields 
\begin{equation}
\label {eq:Clausius}
\big< \exp \big( - \frac {T \Delta s - q}{k_BT}\big) \big> =1    
\end{equation}
where $\big< \dots \big>$ stands for an average over all possible trajectories. Because $e^x \ge 1+x$, Eq.\eqref{eq:Clausius} predicts
\begin{equation}
\label {eq:Clausius1}
\Delta S = \big < \Delta s \big >  \ge \frac {Q}{T}     
\end{equation}   
where $Q\equiv <q>$ is the average heat. 

Eq.\eqref{eq:Clausius1} provides a microscopic interpretation of the Clausius inequality, i.e., the microscopic heat is intrinsically affiliated with a dynamic process that is subject to fluctuations. If the system is isolated, the Liouville theorem predicts that the probability densities of forward and reverse trajectories are identical. In that case, $\Delta_h s=0$, and the entropy change is the same as production, which is nonnegative according to Eq.\eqref{eq:Sgen2}. Meanwhile, the fluctuation theorem predicts that a negative microscopic entropy production is possible for small systems, implying the likelihood of spontaneous heat transfer from cold to hot from a microscopic perspective.    

The microscopic interpretation of heat provides a convenient starting point to derive the work and free energy relations. As a system evolves along a trajectory $\pmb{\nu}(\cdot)$, the conservation of energy requires
\begin{equation}
\label {eq:Work}
\Delta \epsilon = w  + q      
\end{equation}   
where $\Delta \epsilon \equiv \epsilon_\tau - \epsilon_0$ represents the change in the system energy, $w$ and $q$ are work and heat affiliated with trajectory $\pmb{\nu}(\cdot)$, respectively. If the system is in equilibrium at the initial and final states, the Boltzmann equation can be utilized to describe their microstate distributions 
\begin{equation}
\label {eq:BoltzFluc}
p(\pmb{\nu}_t) = \exp \big[ -\frac{\epsilon_{\pmb{\nu}_t}-F_t}{k_BT} \big]     
\end{equation}   
where subscribe $t$ stands for the terminal microstates at $t=0$ and $t=\tau$, and $F_t$ for the corresponding free energy. According to Eq.\eqref{eq:twoDs}, the change in microscopic entropy can be expressed as
\begin{equation}
\label{eq:wfluc0}
\Delta s= k_B \ln \frac{p(\pmb{\nu}_0, 0)}{ p(\pmb{\nu}_\tau, \tau)} = \frac{\Delta \epsilon-\Delta F}{T} =\sigma +q/T
\end{equation}
where $\Delta F \equiv F_\tau -F_0$. Substituting $q = \Delta \epsilon -w $ into Eq.\eqref{eq:wfluc0} yields 
\begin{equation}
\label{eq:wfluc1}
\sigma=\frac{w-\Delta F}{T}.
\end{equation}
Because both $\Delta F$ and $T$ are fixed by the initial and final conditions, substituting Eq.\eqref{eq:wfluc1} into Eqs.\eqref{eq:Sfluc} and \eqref{eq:Sfluc0} predicts 
\begin{equation}
\label{eq:Jarzyn}
\big< \exp \big(- \frac{w}{k_BT}\big)  \big> = \exp \big(-\frac {\Delta F}{k_BT} \big) \\
\end{equation} 
\begin{equation}
\label{eq:Crooks}
\frac{p(w)}{\widetilde{p}(-w) } = \exp \big( \frac{w-\Delta F}{k_BT} \big)
\end{equation} 
Eqs.\eqref{eq:Jarzyn} and \eqref{eq:Crooks} are known as the Jarzynski equality \cite{RN11} and the Crooks fluctuation theorem\cite{RN12}, respectively.  

\textit{Conclusions:} Often regarded as phenomenological, thermodynamics laws are universally applicable to any systems regardless of size or microscopic detail. Here we provide a deductive proof of the second law of thermodynamics that is not limited by any particular setting or assumption. Whereas many equations have been derived before in different contexts with various conjectures (e.g., assuming Markovian dynamics\cite{van2015ensemble}), the derivations shown above are more general and thus more fundamental than existing results. 

We demonstrate that, at any moment, the microscopic entropy can be defined by the probability of the system in specific microstates as suggested by the information theory and the Gibbs-Shannon entropy. An ensemble average of the microscopic entropy is identical to that averaging over all possible trajectories. Therefore, the change in microscopic entropy can be attributed to the stochastic nature of the dynamic process underlying the particle motions, and to the statistics of trajectories inherent to thermodynamic systems. The two distinct sources of uncertainties define the microscopic entropy flow and the microscopic entropy production. The microscopic entropy production is never negative on average and varies with different trajectories as predicted by the fluctuation theorem. 

When the system is in contact with a thermal bath, the flow of microscopic entropy defines the heat transfer between the system and the surrounding. Like entropy production, the Clausius inequality is valid from a statistical perspective, and the fluctuation theorem implies that spontaneous heat transfer from cold to hot is not prohibited from a microscopic perspective. Finally, we demonstrate that the Jarzynski/Crooks equations can be readily obtained from microscopic heat, circumventing any specifications on the microscopic work and the exact forms of particle motions. The distinction between statistical and dynamic uncertainties is applicable to systems of any size, quantum or classical. It can be utilized to define thermodynamic efficiency and, eventually, predict nonequilibrium phenomena based on the motions of individual particles. \\

{\bf{ACKNOWLEDGMENTS}}
The author wishes to acknowledge financial support by the the National Science Foundation’s Harnessing the Data Revolution (HDR) Big Ideas Program (Grant No. NSF 1940118) and by the Fluid Interface Reactions, Structures, and Transport (FIRST) Center, an Energy Frontier Research Center funded by the U.S. Department of Energy (DOE), Office of Science, Office of Basic Energy Sciences.\\ 

{\bf{DATA AVAILABILITY}}
The data that support the findings of this study are available from the corresponding author upon reasonable request.\\

{\bf{REFERENCES}}
\bibliographystyle{aipnum4-1} 
\bibliography{PRL_Thermo}
 
\end{document}